\pdfoutput=1
\documentclass{PoS}

\usepackage{cite}
\usepackage{mciteplus}

\newcommand{\pt}{\ensuremath{p_\mathrm T}}
\newcommand{\RAA}{\ensuremath{R_\mathrm{AA}}}
\newcommand{\sNN}{\ensuremath{s_\mathrm{NN}}}

\title{High-$\pt$ processes measured with ALICE at the LHC}

\ShortTitle{High-$p_{T}$ processes}

\author{\speaker{Jacek Otwinowski}\thanks{On behalf of the ALICE Collaboration}\\
        GSI Helmholtz Centre for Heavy Ion Research, Darmstadt, Germany \\
        E-mail: \email{j.otwinowski@gsi.de}}

\abstract{The study of single-particle and jet production in heavy-ion collisions provides insights into the density of the medium and the energy-loss mechanisms. The observed suppression of high-$\pt$ particle production is generally attributed to energy loss of partons as they propagate through the hot and dense QCD medium - Quark-Gluon-Plasma (QGP). Such measurements allow the characterization of the QGP, the deconfined state of quarks and gluons, predicted by QCD. In these proceedings we present the analysis results of Pb--Pb collisions at $\sqrt{\sNN}=2.76$~TeV recorded by ALICE. The nuclear modification factors ($\RAA$) and the results from jet reconstruction in Pb--Pb are presented. Comparison with other measurements and with theory models is discussed.}

\FullConference{Xth Quark Confinement and the Hadron Spectrum,\\
		October 8-12, 2012\\
		TUM Campus Garching, Munich, Germany}

\begin{document}

\section{Introduction}

This paper reports on measurements of single-particle spectra and jets as a function of transverse momentum ($\pt$) and event centrality in Pb--Pb collisions at $\sqrt{\sNN}=2.76$~TeV recorded  by ALICE  {\cite{ALICE_DET}} in the fall of 2010 and 2011. 

Previous results from the Relativistic Heavy Ion Collider (RHIC) \cite{BRAHMS_WHITE, PHOBOS_WHITE, STAR_WHITE, PHENIX_WHITE, SPS, PHENIXRAA130, STARRAA130, STARRAA, PHENIXRAA, PHENIXPI0RAA} showed that hadron production at high $\pt$ in central Au--Au collisions at $\sqrt{\sNN}=200$~GeV is suppressed by a factor 4--5 compared to expectations from an independent superposition of nucleon-nucleon collisions. This observation is typically expressed in terms of the nuclear modification factor

\begin{equation}\label{EQUATION_RAA}
R_{\rm{AA}} (p_{\rm T} ) = \frac{{\rm d}^{2}N^{\rm{AA}}_{\rm{ch}}/{\rm d}\eta{\rm d}p_{\rm{T}} }{\langle T_{\rm {AA}} \rangle {\rm d}^{2}\sigma^{\rm{pp}}_{ch}/{\rm d}\eta {\rm d}p_{\rm T} }
\end{equation}
where $N^{\rm{AA}}_{\rm{ch}}$ and $\sigma^{\rm{pp}}_{\rm{ch}}$ represent the charged particle yield in nucleus-nucleus (AA) collisions and the cross section in pp collisions, respectively. The nuclear overlap function $T_{\rm{AA}}$ is calculated from the Glauber model \cite{MILLER_TAA} and averaged over each centrality interval, $\langle T_{\rm{AA}}  \rangle = \langle N_{\rm{coll}}  \rangle / \sigma^{\rm{NN}}_{\rm{inel}}$, where $\langle N_{\rm{coll}} \rangle$ is the average number of binary nucleon-nucleon collisions and $\sigma^{\rm{NN}}_{\rm{inel}}$ is the inelastic nucleon-nucleon cross section. In absence of nuclear modifications $\RAA$ is equal to unity at high $\pt$.

\section{$\RAA$ of charged particles}

\begin{figure}[ht]
\begin{center}
\includegraphics[width=0.8\textwidth]{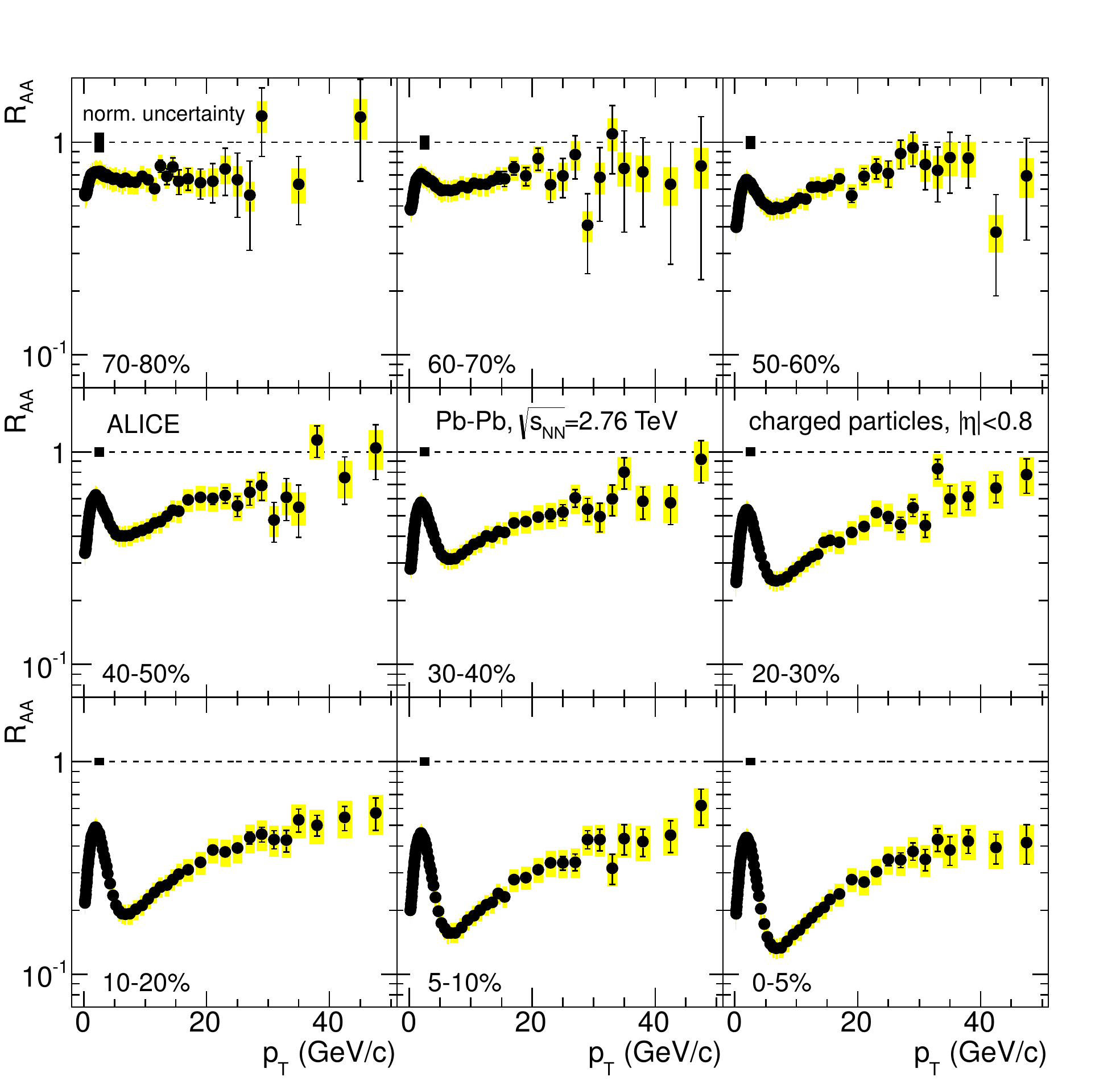}
\caption{$\RAA$ for charged particles as a function of $\pt$ in nine centrality intervals \cite{ALICE_RAA}. Normalization uncertainties are shown as the boxes at $\RAA=1$.}
\label{figRAA}
\end{center}
\end{figure}

\begin{figure}[t]
\begin{center}$
\begin{array}{cc}
\includegraphics[width=2.9in]{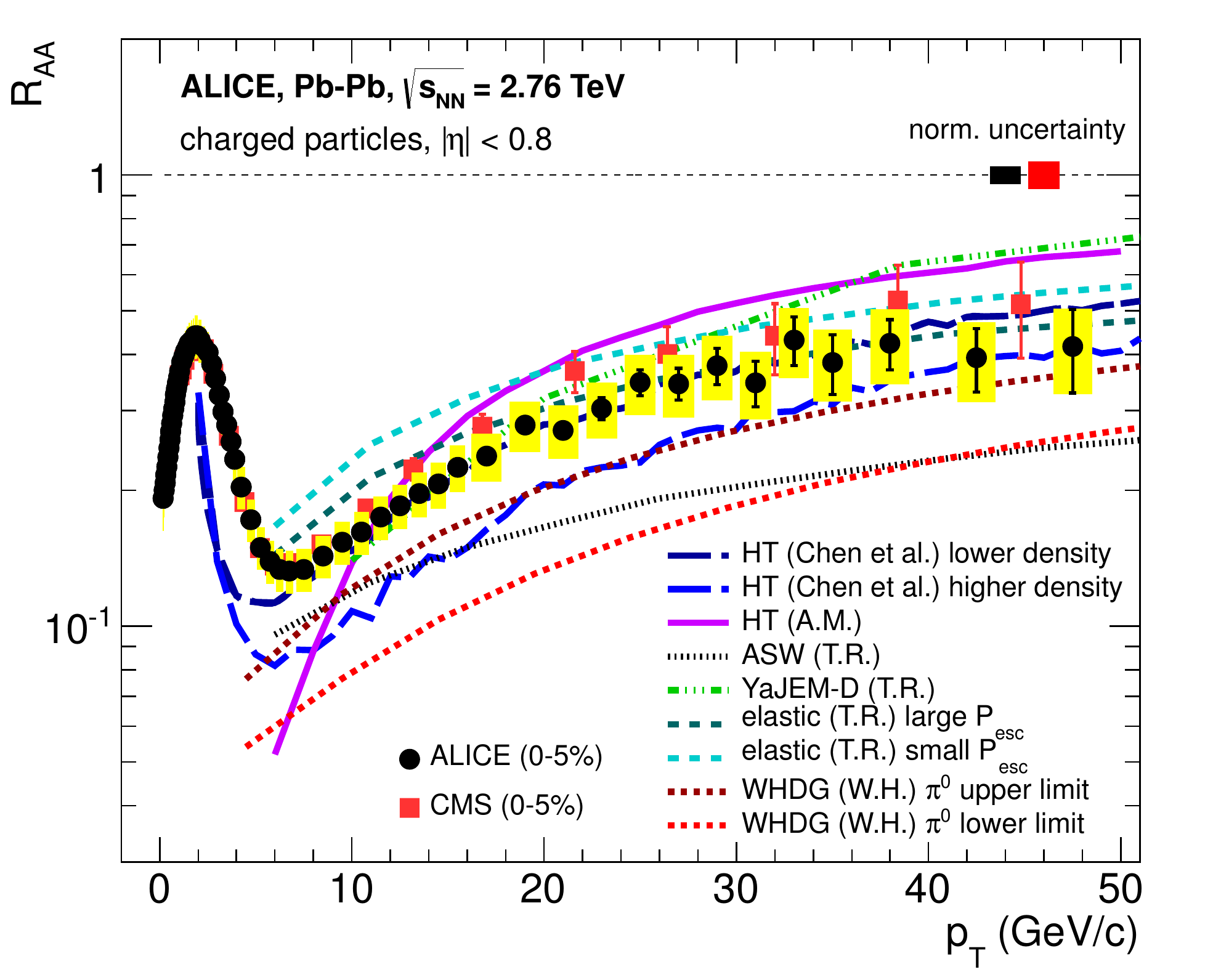} &
\includegraphics[width=2.9in]{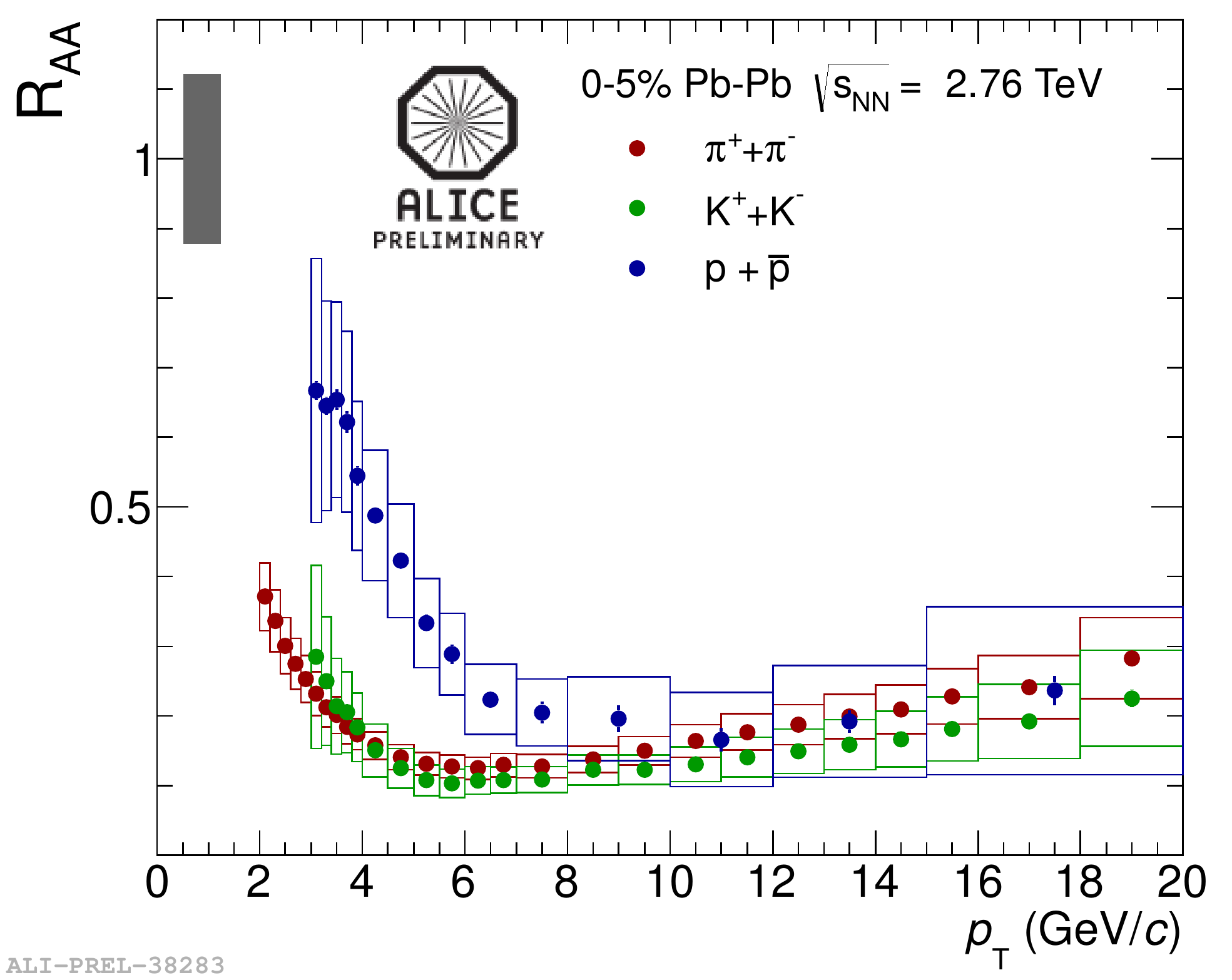}
\end{array}$
\caption{{\bf Left:} $\RAA$ for charged particles as a function of $\pt$ in central Pb--Pb collisions \cite{ALICE_RAA} in comparison to CMS result \cite{CMS_RAA} and model predictions \cite{HOROWITZ2011, ASW, CHEN2011, MAJUMDER2011, RENK2011, RENK2011YAJEM}. { \bf Right:} $\RAA$ for charged pions, kaons and protons as a function of $\pt$ in central Pb--Pb collisions. {\bf Both:} Normalization uncertainties are shown as the boxes at $\RAA=1$.}
\label{figRAAIdent}
\end{center}
\end{figure}

The nuclear modification factors as a function of $\pt$ for nine centrality intervals \cite{ALICE_RAA} are shown in  Fig.\ref{figRAA}. In peripheral collisions (70--80\%), only moderate suppression ($R_{\rm{AA}}=0.6$--0.7) and a weak $\pt$ dependence are observed. Towards more central collisions, a pronounced minimum at about $p_{\rm T}=6$--7~GeV/$c$ develops while for \mbox{$p_{\rm T}>7$~GeV/$c$} there is a significant rise of the nuclear modification factor.
This rise becomes gradually less steep with increasing $p_{\rm T}$. In the most central collisions (0--5\%), the yield is most
suppressed, $R_{\rm{AA}} \approx0.13$ at $p_{\rm T}=6$--7~GeV/$c$, and $R_{\rm{AA}}$ reaches $\approx$~0.4 with no significant $\pt$ dependence for  $p_{\rm T}>30$~GeV/$c$.

The ALICE measurement of $R_{\rm{AA}}$ in the most central \mbox{Pb--Pb} collisions (0--5\%) \cite{ALICE_RAA} is compared to the CMS result \cite{CMS_RAA} in Fig.~\ref{figRAAIdent} (left).  Both measurements agree within their respective statistical and systematic uncertainties. In Fig.~\ref{figRAAIdent} (left), the measured $R_{\rm{AA}}$ for 0--5\% central collisions is also
compared to model predictions. All selected models use RHIC data to calibrate the medium density. All model calculations, except WHDG \cite{HOROWITZ2011}, use a hydrodynamical description of the
medium, but different extrapolation assumptions from RHIC to
LHC. A variety of energy loss formalisms is used. An increase of $R_{\rm{AA}}$ due to a decrease of the relative
energy loss with increasing $p_{\rm T}$  is seen for all the models. The curves labeled WHDG, ASW, and Higher Twist (HT) are based on analytical radiative
energy loss formulations that include interference effects. Of those
curves, the multiple soft gluon approximation (ASW \cite{ASW}) and the
opacity expansion (WHDG \cite{HOROWITZ2011}) show a larger suppression
than seen in the measurement, while one of the HT curves
(Chen \cite{CHEN2011}) with lower parton density provides a good description. The other HT
(Majumder \cite{MAJUMDER2011}) curve shows a stronger rise with
$p_{\rm T}$ than measured. The elastic energy loss model by Renk (elastic)
\cite{RENK2011} does not rise steeply enough with $p_{\rm T}$
and overshoots the data at low $p_{\rm T}$. The YaJEM-D model \cite{RENK2011YAJEM}, which is
based on medium-induced virtuality increases in a parton shower, shows
too strong a $p_{\rm T}$-dependence of $R_{\rm{AA}}$ due to
a formation time cut-off.

A more systematic study of the energy loss formalisms, preferably with the same model(s) for the
medium density is needed to rule out or confirm the various effects. Deviations of the nuclear parton
distribution functions (PDFs) from a simple scaling of the nucleon PDF with mass number A (e.g. shadowing) are also expected to affect the nuclear modification factor. These effects are predicted \cite {SHADOW} to be small for $\pt>10$~GeV/$c$ at the LHC and will be quantified in future p--Pb measurements.

\section{$\RAA$ of identified hadrons}

The measurement of pions, kaons and protons allows us to study the suppression pattern for mesons and baryons, which gives a handle on how to separate quark and gluon energy loss \cite{PARTON}. The analysis of charged pions, kaons and protons at high $\pt$ is based on statistical particle identification using the
specific energy loss d$E$/d$x$ in the TPC \cite{TPCdEdx}. In the region of the relativistic rise of the energy
loss ($\pt > 3$~GeV/$c$), the separation of pions from kaons and protons is nearly independent
of $\pt$ out to $\pt = 50$~GeV/$c$. The fraction of pions, kaons and protons from all charged particles is determined
in bins of $\pt$ by fitting the d$E$/d$x$ distribution with four Gaussians for electrons, pions, kaons and protons. The $\RAA$ for charged pions, kaons and protons identified in central Pb--Pb collisions are shown in Fig.~\ref{figRAAIdent} (right). Production of pions, kaons and protons is strongly suppressed in central Pb--Pb collisions. The yields are most suppressed at $\pt=$~6--8~GeV/$c$, $\RAA\simeq$~0.1--0.2 for all particle species as expected from the $\RAA$ of charged particles shown in Fig.~\ref{figRAAIdent} (left).  For $\pt<7$~GeV/$c$, the $\RAA$ for pions is similar to that for kaons and is smaller than for protons, which is in line with an enhanced baryon production observed in central Pb--Pb collisions \cite{BaryonProd}. For  $\pt>7$~GeV/$c$, the $\RAA$ for pions, kaons and protons approach the same value. 
 
\begin{figure}[t]
\begin{center}$
\begin{array}{cc}
\includegraphics[width=2.8in]{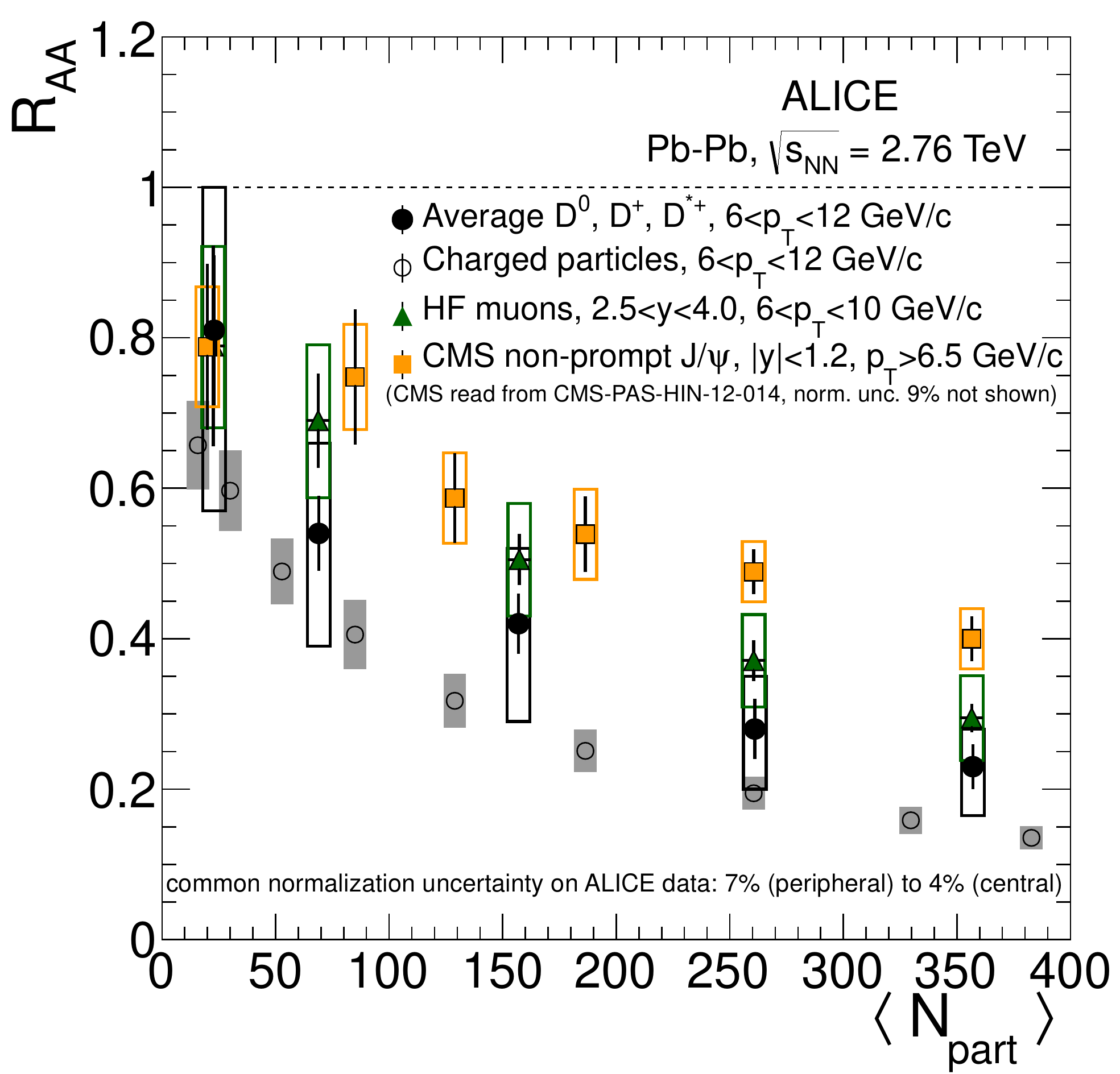} &
\includegraphics[width=3.3in]{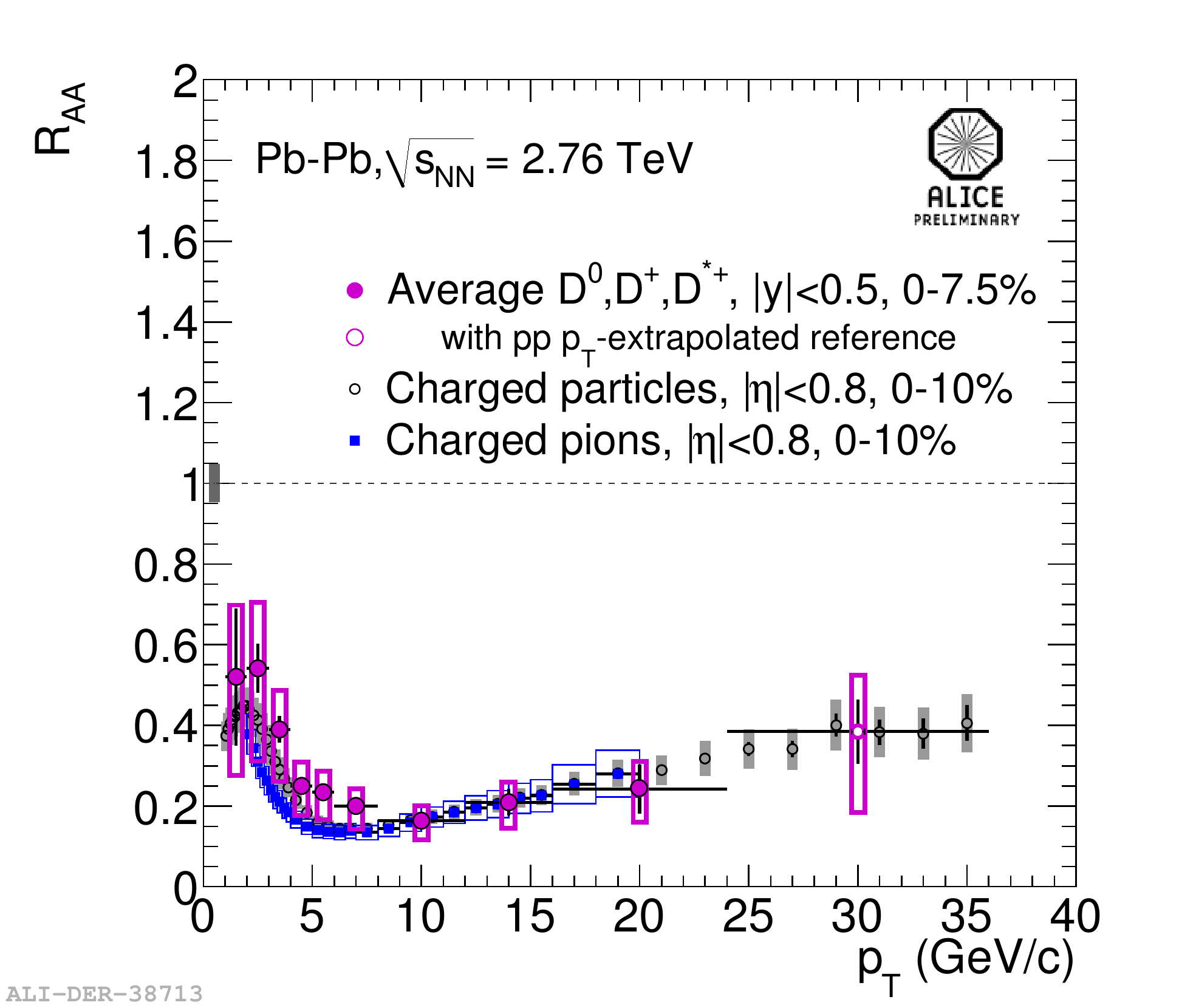}
\end{array}$
\caption{{\bf Left:} $\RAA$ for D mesons \cite{ALICE_RAA_D_mesons}, heavy-flavour decay muons (c,b$\rightarrow\mu$) \cite{ALICE_RAA_muons} and charged particles \cite{ALICE_RAA} as a function of collision centrality in comparison to non-prompt J$/\psi$ from CMS \cite{CMS_RAA_B}. { \bf Right:} $\RAA$ for D mesons, charged particles and charged pions as a function of $\pt$ in central Pb--Pb collisions. Common normalization uncertainties on ALICE data are shown as the box at $\RAA=1$. }
\label{figRAAD}
\end{center}
\end{figure}

The measurement of heavy flavour production at high $\pt$ provides unique observables of jet
quenching. The suggestion that massive quarks experience reduced energy loss due to the suppression of forward radiation ("dead cone effect") has not been borne out by measurements at RHIC.  The $\RAA$ measured for high-$\pt$ D mesons  \cite{ALICE_RAA_D_mesons}, heavy-flavour decay muons (c,b~$\rightarrow\mu$) \cite{ALICE_RAA_muons}  and charged particles  \cite{ALICE_RAA} as a function of collision centrality measured by ALICE are shown in Fig.~\ref{figRAAD} (left). For comparison,  the $\RAA$ for non-prompt J$/\psi$ originating from the B-meson decays measured by CMS \cite{CMS_RAA_B} is also shown. Heavy flavour production at high $\pt$ ($\pt>6$~GeV/$c$) is strongly suppressed in central Pb--Pb collisions, $\RAA\simeq$~0.2--0.4. The suppression pattern is similar for heavy flavour measured by ALICE in central and forward rapidities. The data indicate that there might be a hierarchy in the suppression of the light and heavy flavour, $\RAA^{\rm{B}}>\RAA^{\rm{D}}>\RAA^{\rm{light~flavour}}$. The D-meson $\RAA$ measured as a function of $\pt$ in the most central Pb--Pb collisions is compared to the $\RAA$ for charged particles and pions in Fig.~\ref{figRAAD} (right). Data show that the $\RAA$ of D mesons is consistent with the $\RAA$ of light-flavour hadrons at high $\pt$ ($>8$~GeV/$c$). 
Higher statistics of Pb--Pb data and comparison data in p--Pb collisions should allow us to study this region with more precision and disentangle initial-state nuclear effects, which could be different for light and heavy flavour.
 
 \section{$\RAA$ of charged jets}
 
We present results for the jets reconstructed from the charged particles in Pb--Pb collisions measured in 2010. The jet reconstruction with charged and neutral particles (2011 run) is being finalized. The particle tracks are reconstructed using the Time Projection Chamber (TPC) and vertexing
information from the Inner Tracking System (ITS). This ensures maximum azimuthal angle ($\phi$) uniformity of reconstructed tracks with transverse momenta down to $\pt=150$~MeV/$c$.

The jets are reconstructed using the anti-$k_{\rm{T}}$ algorithm \cite{ANTIKT} and are corrected for the background in each event using the jet area $A$ with $p^{\rm{ch}}_{\rm{T,jet}}  = p^{\rm{rec}}_{\rm{T,jet}} - \rho \cdot A$. Here, the background density $\rho$ is calculated on an event-by-event basis as the median for the $p_{\rm{T}}/A$ ratio of reconstructed $k_{\rm{T}}$-clusters in $|\eta|< 0.5$ by using $k_{\rm{T}}$ algorithm \cite{KT}. In addition, the charged jets are corrected for the background fluctuations using an unfolding procedure \cite{JETSBACKG}.

\begin{figure}[t]
\begin{center}$
\begin{array}{cc}
\includegraphics[width=3.1in]{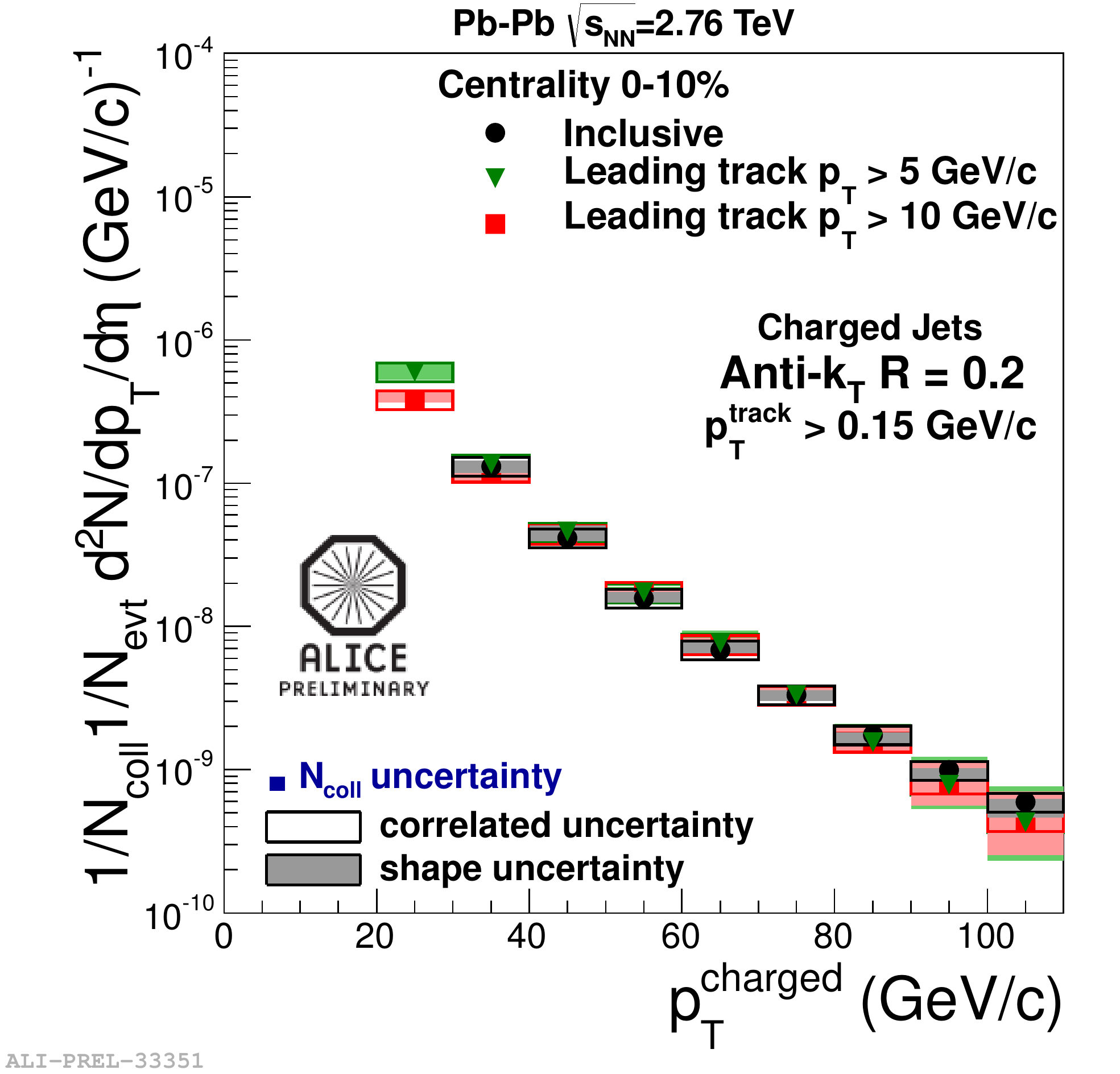} &
\includegraphics[width=2.7in]{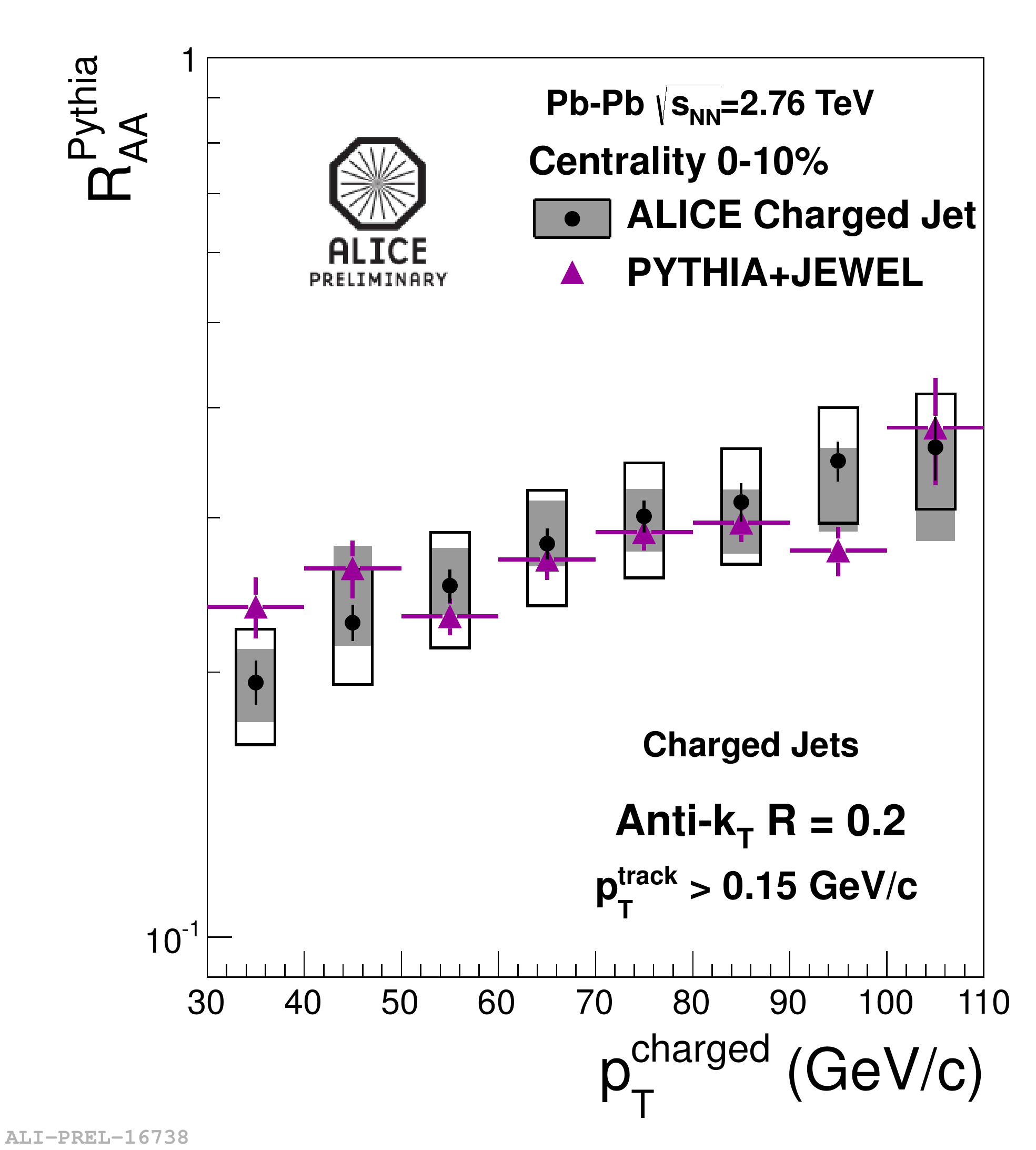}
\end{array}$
\caption{{\bf Left:} $\pt$ spectra for inclusive charged jets and for jets with the threshold on the leading particle $\pt$. { \bf Right:} $\RAA$ for charged jets in central Pb--Pb collisions in comparison to the model calculations \cite{JEWEL}. pp reference spectrum for $\RAA^{\rm{Pythia}}$ based on Pythia \cite{PythiaPerugia}.}
\label{figRAAJets}
\end{center}
\end{figure}

The $\pt$ spectra for jets reconstructed in Pb--Pb collisions from charged
particles with $\pt>0.15$ and jet radius R = 0.2 are shown in Fig.~\ref{figRAAJets} (left). The measurement was done for inclusive (unbiased) jets and for jets with  the threshold on the leading particle $\pt$ requirement ($\pt>5$~GeV/$c$ and $\pt>10$~ GeV/$c$). The $\pt$ spectra for inclusive jets are similar to those for the jets with the leading particle selection indicating that the unfolding procedure deals properly with the background at low $\pt$ (large background fluctuations) and there is rather a weak softening of the fragmentation in central Pb--Pb collisions.

The $\RAA^{\rm{Pythia}}$ for charged jets as a function of $\pt$ in  central Pb-Pb collisions in comparison to model calculations (JEWEL \cite{JEWEL}) is shown in Fig.~\ref{figRAAJets} (right). The $\RAA^{\rm{Pythia}}$ was constructed using the $\pt$ spectrum for charged jets from Pythia \cite{PythiaPerugia} as a pp reference.  A strong charged jet suppression is observed with $\RAA\simeq$~0.2--0.4 in central Pb--Pb collisions which is consistent with the $\RAA$ measured for the charged particles \cite{ALICE_RAA}, and is in agreement with the ATLAS \cite{ATLAS_Jet_RAA} and CMS \cite{CMS_Jet_RAA} results. A good agreement between the JEWEL model calculations and data is observed. 

\section{Summary}

The results indicate a strong suppression of charged particle production
in Pb--Pb collisions at $\sqrt{\sNN}=2.76$~TeV and a characteristic centrality and $\pt$
dependence of the nuclear modification factors. The data indicate that there might be a hierarchy in the suppression of the light and heavy flavour at high $\pt$ (> 6~GeV/$c$), $\RAA^{\rm{B}}>\RAA^{\rm{D}}>\RAA^{\rm{light~flavour}}$. The nuclear modification factor for charged jets is similar to that for charged particles, and is in agreement with the ATLAS and CMS results. The comparison of ALICE data to model calculations indicates a large sensitivity of high-$\pt$ particle production to details of energy loss mechanisms.

\end{document}